\documentclass[a4,aps,pre,twocolumn,10pt,superscriptaddress]{revtex4-1}
\usepackage{amsmath}
\usepackage{amsfonts}
\usepackage{amssymb,enumerate}
\usepackage{graphicx,hyperref}
\usepackage{amssymb,float}
\usepackage{epstopdf,booktabs}
\usepackage{mathptmx}
\usepackage[usenames,dvipsnames]{color}
\usepackage[T1]{fontenc}
\usepackage[utf8]{inputenc}
\usepackage{natbib}

\newcommand{\be}{\begin {equation}}
\newcommand{\ee}{\end {equation}}
\newcommand{\beqa}{\begin {eqnarray}}
\newcommand{\eeqa}{\end {eqnarray}}

\hypersetup{
    colorlinks,    citecolor=blue,    filecolor=blue,    linkcolor=blue,    urlcolor=blue
}

\begin{document}

\title{Transition from wakefield generation to soliton formation}

\begin{abstract}
It is well known that when a short laser pulse propagates in an underdense
plasma, it induces longitudinal plasma oscillations at the plasma frequency
after the pulse, typically referred to as the \textit{wakefield}. However,
for plasma densities approaching the critical density wakefield generation
is suppressed, and instead the EM-pulse undergoes nonlinear self-modulation.
In this article we have studied the transition from the wakefield generation
to formation of quasi-solitons as the plasma density is increased. For this
purpose we have applied a one dimensional (1D) relativistic cold fluid model, which has also
been compared with particle-in-cell simulations. A key result is that the
energy loss of the EM-pulse due to wakefield generation has its maximum for
a plasma density of the order 10 percent of the critical density, but that
wakefield generation is sharply suppressed when the density is increased
further.
\end{abstract}

 \author{Amol R. Holkundkar}
 \email{amol.holkundkar@pilani.bits-pilani.ac.in}
 \affiliation{Department of Physics, Birla Institute of Technology and Science - Pilani, 
 Rajasthan, 333031, India}

 \author{Gert Brodin}
 \email{gert.brodin@physics.umu.se}
 \affiliation{Department of Physics, Ume\aa\ University, SE - 90187 Ume\aa, Sweden  }

 \date{\today}
 \maketitle

\section{Introduction}

Wakefield generation is of fundamental interest in plasmas, both from a
basic science point of view and when it comes to applications. Of particular
interest is the laser wakefield acceleration scheme, which has shown
tremendous progress since the pioneering work by Tajima and Dawson \cite%
{Tajima1979_prl}. There are numerous experimental demonstration of electron
acceleration to GeV of energies \cite%
{Kneip2009_prl,kim2013_prl,Leemans2014_prl}. Apart from this, a number of
different approaches for electron acceleration have been proposed and demonstrated experimentally by
different research groups around the globe e.g. bubble regime \cite%
{PhysRevE.76.055402,PhysRevLett.104.195002}, beat wave acceleration \cite%
{PhysRevE.69.026404,PhysRevE.79.056406}, self modulated laser wakefield
acceleration \cite{PhysRevE.73.016405}, and many more. Furthermore, the
effects of external magnetic fields \cite{PhysRevE.84.036409}, effects of
modifying the laser chirp \cite{1367-2630-14-2-023057}, and effects of
varying the plasma density profile \cite{PhysRevE.77.025401} on wakefield
generation, have drawn considerable research interest and continue to do so.

When the electromagnetic pulses are long (as compared to the skin depth),
wake field generation is suppressed, and instead other nonlinear phenomena
becomes more pronounced. Typically in a non-magnetized plasma the
nonlinearity is of a focusing type, which can allow for envelope bright
solitons \cite{Yu1982_PF}. Soliton formation have different features
depending on the dimensionality, and lot of interest has been devoted to the
2D \cite{NaumovaPRL_2001} and 3D phenomena \cite{EsirkepovPRL_2002}.
However, when the pulses are pancake shaped, the physical scenario can be
described by a 1D-model \cite{Jovanovic2015_POP} to a good approximation.

In the present paper we will study the competing mechanisms of linear
dispersion, soliton formation and wakefield generation, and their dependence
on wave amplitude and plasma density. For this purpose we apply an 1D
relativistic cold fluid model. One of the main results in the present paper
is that the peak in wakefield energy density occurs for densities around $%
n\approx 0.1n_{c}$, where $n$ is the plasma density and $n_{c}$ is the
critical density, but this is followed by a very sharp decrease in the wake
field energy for densities $n\gtrsim 0.2n_{c}$. Furthermore, the effect of
varying the laser amplitude is studied. The results from the cold
relativistic fluid model are compared with particle-in-cell (PIC)
simulations, and the agreement is found to be excellent.

The organization of our paper is as follows. In section II the governing
equations for 1D wave propagation are derived (based on a cold relativistic
fluid model). Next in section III the basic equations are solved
numerically, and our main results are presented. In section IV we make a
comparison with a simplified theory and with 1D PIC simulations.  Finally in
Section V we make a summary and present the final conclusions.

\section{The cold relativistic fluid model}

The main purpose of the present article is to study the propagation of an 1D
electromagnetic pulse based on the cold relativistic fluid equations for
electrons. Apart from treating the ions as immobile, no further
approximations are made, and the equations will be solved numerically. The
numerical results will be described in the next section. Before implementing
the scheme, we first demonstrate that the number of basic equations needed
can be reduced somewhat.

We assume all fields to depend on $(z,t)$ and divide the fields in the
parallel and perpendicular direction to $z$ (e.g. the velocity field is
written as $\mathbf{v}=\mathbf{v}_{||}+\mathbf{v}_{\perp }$ with $\mathbf{%
v_{z}}\equiv \mathbf{v}_{||}$). Using the Coulomb gauge, we immediately
obtain from Gauss' law and the perpendicular component of Ampere's laws 
\begin{equation}
\frac{\partial ^{2}\phi }{\partial z^{2}}=\frac{e(n_{e}-n_{0})}{\varepsilon
_{0}}
\end{equation}%
and%
\begin{equation}
\nabla ^{2}\mathbf{A}_{\perp }-\frac{1}{c^{2}}\frac{\partial ^{2}\mathbf{A}%
_{\perp }}{\partial t^{2}}=\mu _{0}n_{e}\mathbf{v}_{\perp }.  \label{Amp-1}
\end{equation}%
Here $e$ is the elementary charge, $n_{e}$ is the electron number density
and $n_{0}$ represents the constant neutralizing ion background, $\phi $ is
the scalar potential and $\mathbf{A}_{\perp }$ is the perpendicular part of
the vector potential. Noting that the Coulomb gauge and the 1D geometry
implies $A_{z}=0$, the parallel component of Ampere's law is written 
\begin{equation}
\frac{1}{c^{2}}\frac{\partial ^{2}\phi }{\partial t\partial z}=-\mu
_{0}en_{e}\upsilon _{z}  \label{Amp-2}
\end{equation}%
For cold electrons, only the Lorentz force is needed in the electron
equation of motion. Dividing the equation in its parallel and perpendicular
components gives%
\begin{equation}
\frac{d\mathbf{P_{\perp }}}{dt}=e\Big[\frac{\partial }{\partial t}+\upsilon
_{z}\frac{\partial }{\partial z}\Big]\mathbf{A_{\perp }}=\frac{d}{dt}(e%
\mathbf{A_{\perp }})  \label{pperp}
\end{equation}%
and%
\begin{equation}
\frac{d\mathbf{P_{z}}}{dt}=e\frac{\partial \phi }{\partial z}\mathbf{e_{z}}-e%
\Big(\mathbf{v_{\perp }}\cdot \frac{\partial \mathbf{A_{\perp }}}{\partial z}%
\Big)\mathbf{e_{z}}  \label{pz}
\end{equation}%
where $\mathbf{P}$ is the momentum and $d/dt\equiv \partial /\partial
t+\upsilon _{z}\partial /\partial z$ is the total (convective) time
derivative. Eq. (\ref{pperp}) can be integrated to give 
\begin{equation}
\mathbf{P_{\perp }}=e\mathbf{A_{\perp }}\implies \mathbf{v_{\perp }}=e%
\mathbf{A_{\perp }}/\gamma m_{e}  \label{4sol}
\end{equation}%
where $m_{e}$ is the electron mass and $\gamma =\sqrt{1+\mathbf{P}^{2}%
\mathbf{/}m_{e}^{2}c^{2}}$ is the relativistic gamma factor. Given (\ref%
{4sol}) we can write Eq. (\ref{pz}) as

\begin{equation}
\frac{dP_{z}}{dt}=\frac{d}{dt}(\gamma m_{e}\upsilon _{z})=e\frac{\partial
\phi }{\partial z}-\frac{e^{2}}{2\gamma m_{e}}\frac{\partial A_{\perp }^{2}}{%
\partial z}
\end{equation}%
which can be further rewritten as

\begin{equation}
\frac{d\upsilon _{z}}{dt}=\frac{e}{\gamma m_{e}}\frac{\partial \phi }{%
\partial z}-\frac{e^{2}}{2\gamma ^{2}m_{e}^{2}}\frac{\partial A_{\perp }^{2}%
}{\partial z}-\frac{\upsilon_{z}}{\gamma }\frac{d\gamma }{dt}  \label{help-1}
\end{equation}%
Noting that the rate of change of the energy is given by $d\mathcal{E}/dt=-e%
\mathbf{v\cdot E}$, where $\mathcal{E=\gamma }mc^{2}$, we find 
\begin{equation}
\frac{d\gamma }{dt}=\frac{e}{m_{e}c^{2}}\Big(\upsilon _{z}\frac{\partial
\phi }{\partial z}+\frac{e}{2\gamma m_{e}}\frac{\partial A_{\perp }^{2}}{%
\partial t}\Big).  \label{gama0}
\end{equation}%
Combing (\ref{gama0}) and (\ref{help-1}) we then obtain 
\begin{equation}
\frac{d\upsilon _{z}}{dt}=\frac{e}{\gamma m_{e}}\Big(1-\frac{\upsilon
_{z}^{2}}{c^{2}}\Big)\frac{\partial \phi }{\partial z}-\frac{e^{2}}{2\gamma
^{2}m_{e}^{2}}\Big(\frac{\partial A_{\perp }^{2}}{\partial z}+\frac{\upsilon
_{z}}{c^{2}}\frac{\partial A_{\perp }^{2}}{\partial t}\Big)
\end{equation}%
Next we want want to rewrite $\gamma $ as a function of $\upsilon_{z}$ and $%
A_{\perp }^{2}$. Using $\gamma =1/\sqrt{1-(\upsilon_{\perp
}^{2}+\upsilon_{z}^{2})/c^{2}}$ and $\upsilon_{\perp }^{2}=(eA\mathbf{%
_{\perp }}/\gamma m_{e})^{2}$ we deduce the expression%
\begin{equation}
\gamma =\sqrt{\frac{1+(eA_{\perp }/m_{e}c)^{2}}{1-\upsilon _{z}^{2}/c^{2}}}
\label{gamma-f}
\end{equation}
Our system is completed with the electron continuity equation

\begin{equation}
\frac{\partial n_{e}}{\partial t}+\frac{\partial }{\partial z}(n_{e}\upsilon
_{z})=0  \label{cont-f}
\end{equation}%
Together Eqs. (\ref{Amp-1}), (\ref{Amp-2}), (\ref{help-1}), (\ref{gamma-f})
and (\ref{cont-f}) constitute a closed set.

Next we introduce the normalizations $\mathbf{a}=e\mathbf{A_{\perp }}/m_{e}c$
and $\varphi =e\phi /m_{e}c^{2}$ for the vector and scalar potential respectively.
Furthermore, time and space are normalized against the laser frequency and
wave number ($\omega t\rightarrow t$ and $kx\rightarrow x$) respectively.
The parallel velocity is normalized against the speed of light, $\beta
=\upsilon _{z}/c$, and finally the electron density is normalized against
the critical density $n_{c}=\varepsilon _{0}\omega ^{2}m_{e}/e^{2}$ (from
now on $n_{e}$ represent $n_{e}/n_{c}$). By using these normalization our
basic equations are written as:

\begin{equation}
\frac{\partial ^{2}\mathbf{a}}{\partial z^{2}}-\frac{\partial ^{2}\mathbf{a}%
}{\partial t^{2}}=n_{e}\frac{\mathbf{a}}{\gamma }  \label{Norm-1}
\end{equation}%
\begin{equation}
\frac{d\beta}{dt}=\frac{(1-\beta ^{2})}{\gamma }\frac{\partial \varphi }{%
\partial z}-\frac{1}{2\gamma ^{2}}\Big(\frac{\partial \mathbf{a}^{2}}{%
\partial z}+\beta\frac{\partial \mathbf{a}^{2}}{\partial t}\Big)
\label{Norm-2}
\end{equation}%
\begin{equation}
\frac{\partial n_{e}}{\partial t}+\frac{\partial }{\partial z}\Big(%
n_{e}\beta \Big)=0  \label{Norm-3}
\end{equation}%
\begin{equation}
\gamma =\sqrt{\frac{1+a^{2}}{1-\beta ^{2}}}  \label{Norm-4}
\end{equation}%
\begin{equation}
\frac{\partial ^{2}\varphi }{\partial t\partial z}=-n_{e}\beta
\label{Norm-5b}
\end{equation}%
Equations (\ref{Norm-1})-(\ref{Norm-5b}) constitute the basis for the result
presented below.

\begin{figure}[b]
\begin{center}
\includegraphics[totalheight=2.2in]{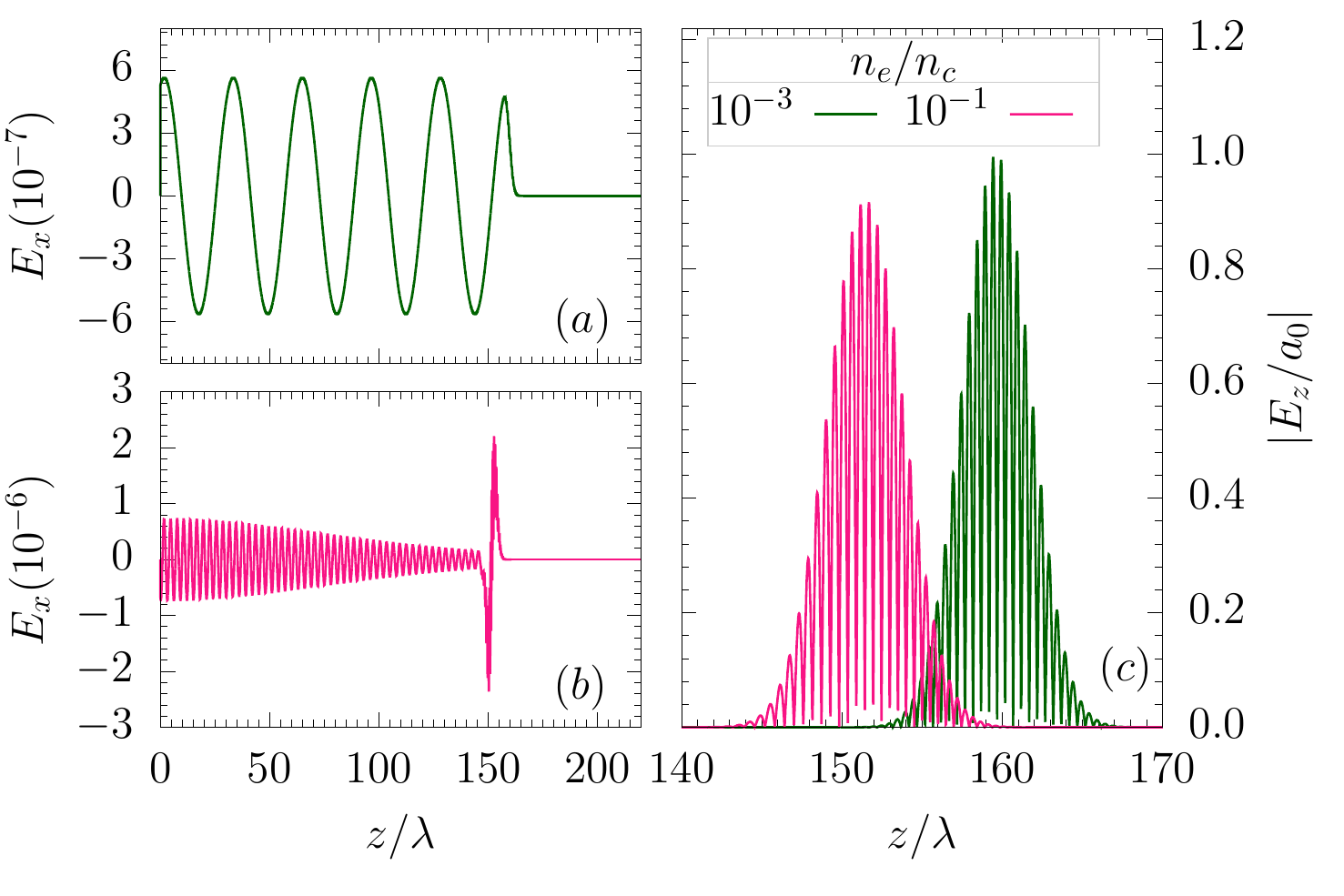}
\end{center}
\caption{The spatial profiles of longitudinal field are plotted at $t = 159$
for the case when a 5 cycle laser pulse with $a_0 = 0.01$ interacts with a 
plasma having density $0.001n_c$ (a) and $0.1n_c$ (b). The corresponding spatial
profiles of the transverse field (EM driver) is also presented in (c). }
\label{lowa0}
\end{figure}

\begin{figure*}[t]
\includegraphics[totalheight=3.6in]{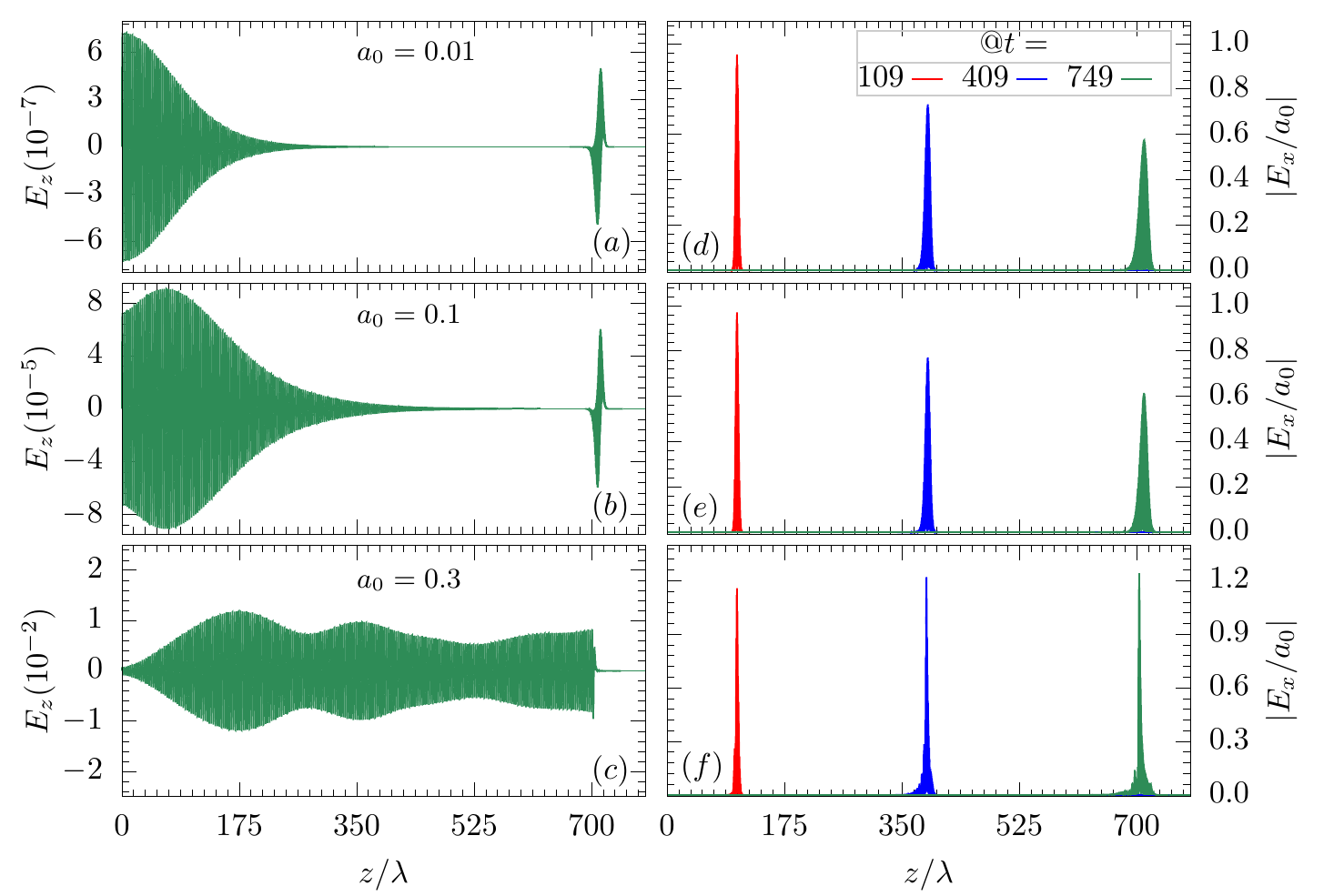}
\caption{The spatial profiles of the longitudinal field (left panel) are plotted
at $t=749$ for the cases when a 5 cycle laser pulse with $a_0 = 0.01$ (a), $%
0.1 $ (b) and $0.3$ (c) interacts with a plasma having density $0.1n_c$. The
spatial profiles of the transverse field  with $a_0 = 0.01$ (d), $%
0.1 $ (e) and $0.3$ (f)  are also presented in right panel for $t
= 109, 409$ and 749. }
\label{dispersion}
\end{figure*}

\section{Results}

We have solved Eqs (\ref{Norm-1})-(\ref{Norm-5b}) numerically in the same
sequence for the case of a localized EM-pulse entering the simulation box
from the left side. The simulation box has been varied in size from $275-750$
$\lambda $ in the various simulations, with a constant unperturbed plasma
density $n_{0}$ throughout the box length. The linearly polarized Gaussian
laser pulse of wavelength $800$ nm has a full width half maximum (FWHM) duration of 5 cycles ($\tau
_{fwhm}=5\times 2\pi $ ). The normalized amplitude $a_{0}$ is varied in the
different simulations, and the pulse is launched from the left boundary of
the simulation box. The boundary conditions on the left boundary read as: 
\begin{equation}
\mathbf{a}(0,t)=a_{0}\exp \Big(-\frac{4\log (2)t^{2}}{\tau _{fwhm}^{2}}\Big)%
\cos (t)\ \ \mathbf{\hat{x}}
\end{equation}%
\begin{equation}
n_{e}(0,t)=n_{0}
\end{equation}%
\begin{equation}
\beta(0,t)=\varphi ^{\prime }(0,t)=0\quad (\varphi ^{\prime }\equiv
\partial \varphi /\partial z).
\end{equation}

In the following we present the results by solving the fluid equations. The
results are categorized in different subsections for pedagogical reasons.

\subsection{The transition from dispersive to non-dispersive pulses}

\begin{figure*}[t]
\begin{center}
\includegraphics[totalheight=3in]{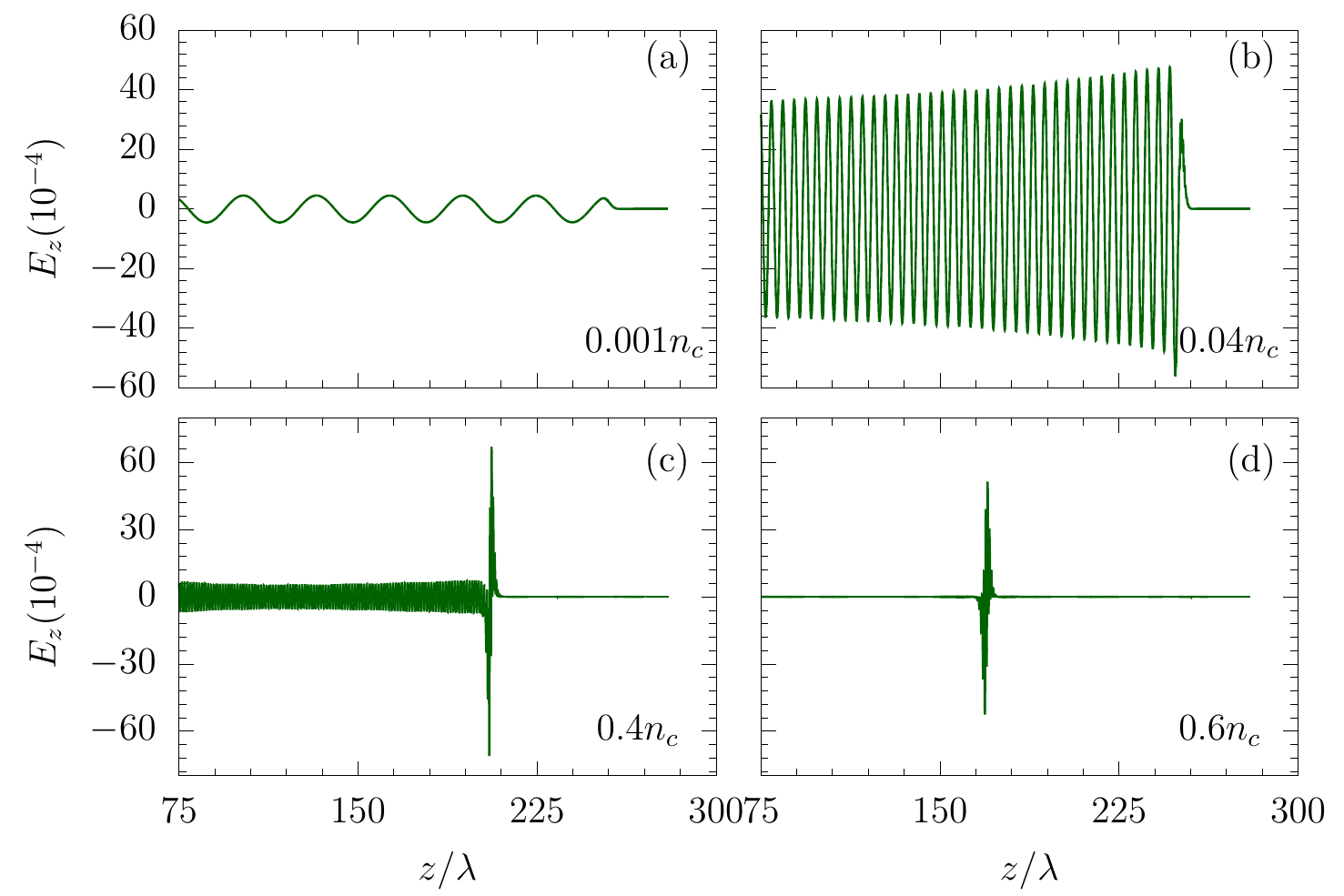}
\end{center}
\caption{Fluid simulation: The wakefield for different plasma densities 0.001%
$n_c$ (a), 0.04$n_c$ (b), 0.4$n_c$ (c) and 0.6$n_c$ (d) are presented as
calculated at $t = 263$. The laser parameters are $a_0 = 0.3$, 5 cycle
(FWHM) Gaussian linearly polarized laser pulse, incident on the plasma slab
of $275\protect\lambda$.}
\label{fluidWake}
\end{figure*}

In Fig. \ref{lowa0}, we have presented the wakefield generation for a low
amplitude laser pulse. The interaction of 5 cycle Gaussian laser pulse with
peak amplitude $a_{0}=0.01$ with an unperturbed plasma with a density of $%
0.001n_{c}$ and $0.1n_{c}$ is considered, and the spatial profiles of
longitudinal (Fig. \ref{lowa0} a,b) and transverse (Fig. \ref{lowa0} c)
fields as evaluated at $t=159$ are shown. 

For the lower density we see that
the longitudinal field has a pure wakefield nature, i.e. harmonic
oscillations are induced after the peak of the EM-pulse. Moreover, since
dispersive effects are very small for such a low density, the wakefield
generation continues more or less unchanged for a long time. By contrast,
for a density of $0.1n_{c}$ the longitudinal field still have a pronounced
wakefield, but there is also a strong peak where the EM-pulse is localized.
Furthermore, for the low amplitude $a_{0}=0.01$ the wake field amplitude is
largest directly after the pulse entrance, and then the wake field amplitude
is gradually decreasing. The reason for the diminishing wakefield amplitude
is the EM-wave dispersion, which become significant for densities of order $%
0.1n_{c}$ and higher. This can be verified by studying Fig \ref{dispersion}.
In Fig \ref{dispersion}(d) we see the EM-wave profile for $\,\,a_{0}=0.01$
at $t=109,409,749$ (note the pulse broadening). As a result of the pulse
broadening, the wakefield generation is decreasing, as can be seen from the
longitudinal field displayed in Fig \ref{dispersion}(a). 

When the amplitude
is increased, the nonlinearity begins to counteract dispersion, which means
that wakefield generation can be sustained. We compare the EM-wave profile
for $\,\,a_{0}=0.01,\,\,a_{0}=0.1$ and $\,\,a_{0}=0.3$ at $t=109,409,749$ in
the right panel of Fig \ref{dispersion}. While there is some tendency to
decreased dispersion for $a_{0}=0.1$, the main suppression of dispersion
occurs when increasing the amplitude up to $a_{0}=0.3$. The result for the
longitudinal fields is quite dramatic, as seen when comparing the
corresponding profiles in the left panel of Fig \ref{dispersion}. Note that
wakefield generation is completely suppressed due to pulse broadening both
for $\,a_{0}=0.01$ and $\,\,a_{0}=0.1$, although the suppression takes
somewhat longer in the latter case. By contrast, the prevention of pulse
broadening for $a_{0}=0.3$ means that wakefield generation can be sustained
for a long time. Naturally, the energy loss due to wakefield generation will
eventually decrease the EM wave amplitude, such that the degree of
nonlinearity is reduced and dispersion sets in, but that will take much
longer time.

\subsection{The transition from wake field generation to soliton formation}

In Fig. \ref{fluidWake}, the spatial profile of longitudinal field $%
E_{z}=-\varphi ^{\prime }$ for different plasma densities calculated at $%
t=263$ ($\sim 42$ cycles) is presented by solving Eqs. (\ref{Norm-1})-(\ref%
{Norm-5b}) numerically for $a_{0}=0.3$. Varying the density from $%
n=0.001n_{c}$ up to $n=0.6n_{c}$ we note that the longitudinal field is a
pronounced wakefield for the lower densities, but gradually turns into a
driven field that is localized to the same region as the EM field that
drives the perturbation. As we increase the initial plasma density, the
wakefield amplitude decreases and eventually a soliton like structure
propagates in the plasma.

\begin{figure}[b]
\begin{center}
\includegraphics[totalheight=2in]{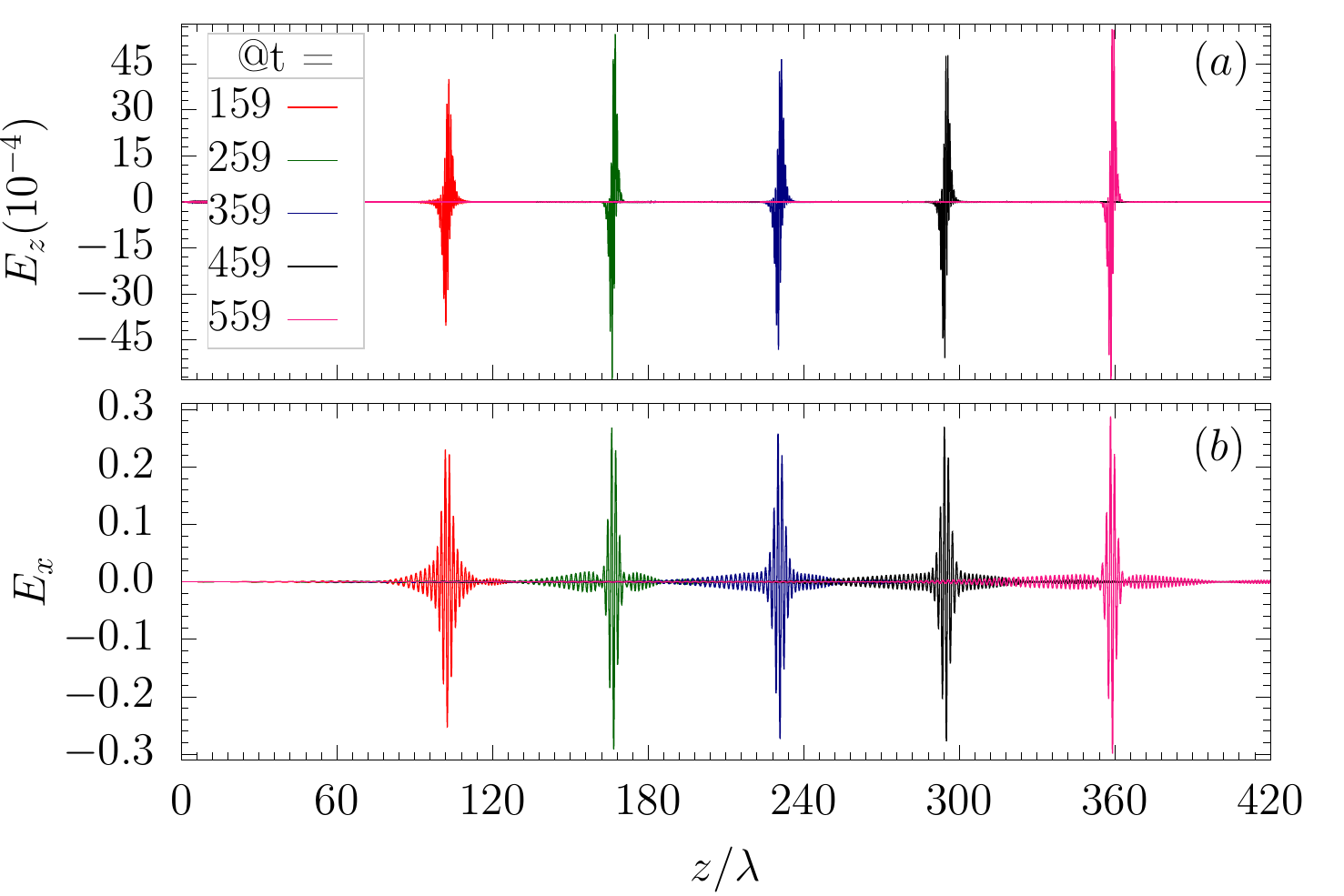} %
\includegraphics[totalheight=2in]{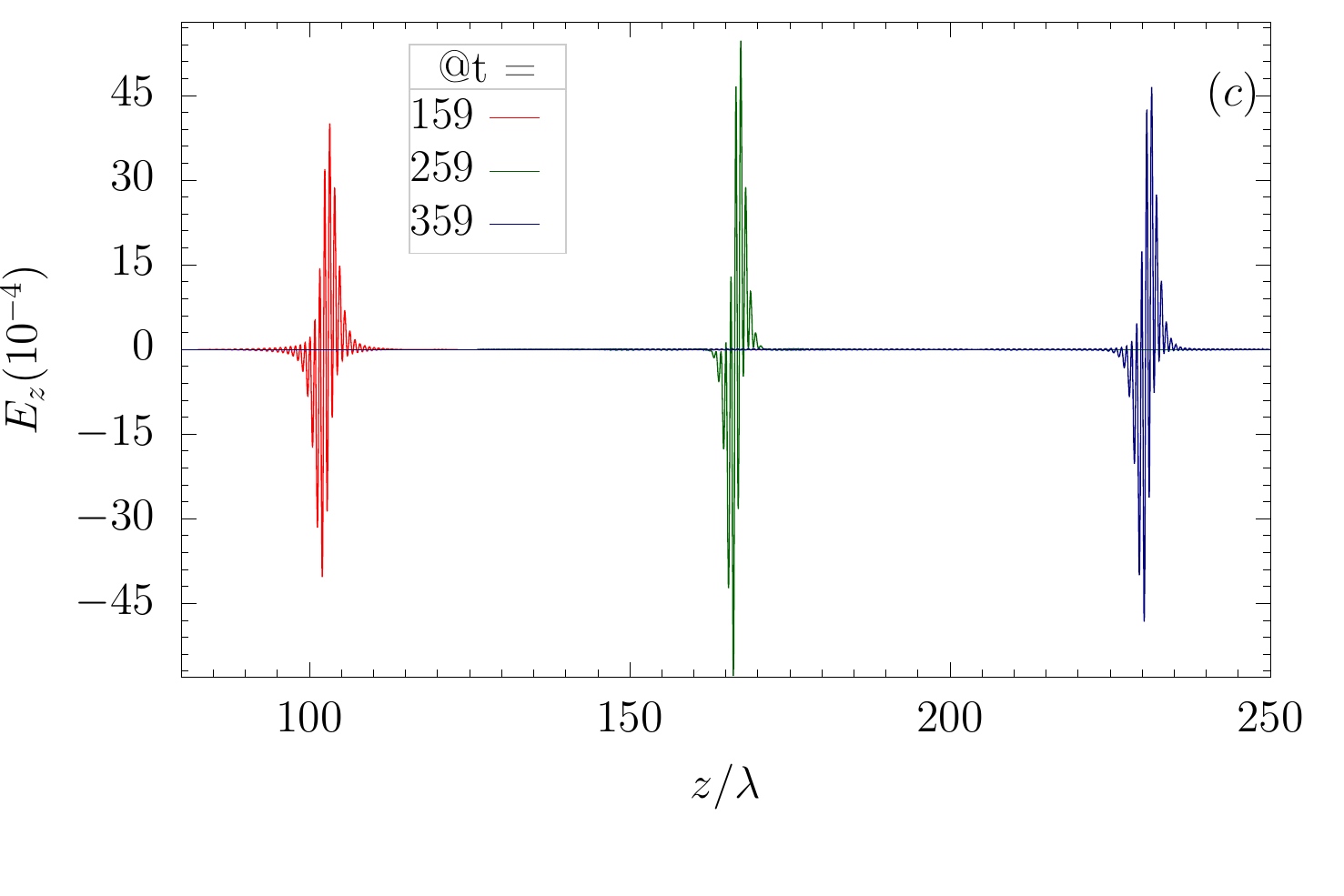}
\end{center}
\caption{Spatiotemporal evolution of electrostatic (a),
electromagnetic (b) fields with 5 cycle laser with amplitude $%
a_0=0.3$ and $n_e = 0.6n_c$. Enlarged version of the solitonic structures is
also presented separately (c). }
\label{solpropa}
\end{figure}

The time evolution of the transverse pulse profile for $a_{0}=0.3$ and $%
n_{e}=0.6n_{c}$ is presented in Fig. \ref{solpropa}(b),
along with the corresponding longitudinal field Fig. \ref{solpropa}(a). As we can
see, the nonlinearity prevents dispersion for most of the energy contained
in the pulse, and the central part of the pulse tends to shorten over time.
As a result the longitudinal field driven by the transverse part tend to
increase somewhat over time. However, we also see that the high- and
low-frequency parts of the pulse spectrum tend to irradiate forwards and
backwards respectively. Apart from this irradiation, which represent a
relatively minor energy loss of the central part of the EM-pulse, the pulse
profile approaches a more or less fixed shape. We will refer to these
structures as \textquotedblleft \textit{quasi-solitons}\textquotedblright .
We cannot exclude that true soliton-formation eventually occur, but for the
rather long time-span that we follow, we see a process where parts of the
frequency spectrum is irradiated backwards and forwards. To give a more
clear view of the longitudinal pulse profile, we present a zoom of the
longitudinal  fields for three different times in Fig. \ref{solpropa}(c).

\subsection{Energy loss as a function of density}

For a density much less than the critical density, wakefield generation is
the dominant process, as seen e.g. in Fig \ref{fluidWake}(a). However, due
to the limited number of particles, the longitudinal field cannot store a
very high energy density, and thus the energy loss for the transverse
degrees of freedom (= the electromagnetic pulse) is limited. For a high
plasma density (i.e. not much smaller than the critical density), wakefield
generation is almost completely suppressed, and there is negligible energy
loss to the longitudinal degrees of freedom, see Fig \ref{fluidWake}(d).
These simple observations imply that there is an intermediate plasma density where
the energy loss due to wakefield generation has its maximum. The purpose of
this subsection is to determine this characteristic value.

\begin{figure}[t]
\begin{center}
\includegraphics[totalheight=2.2in]{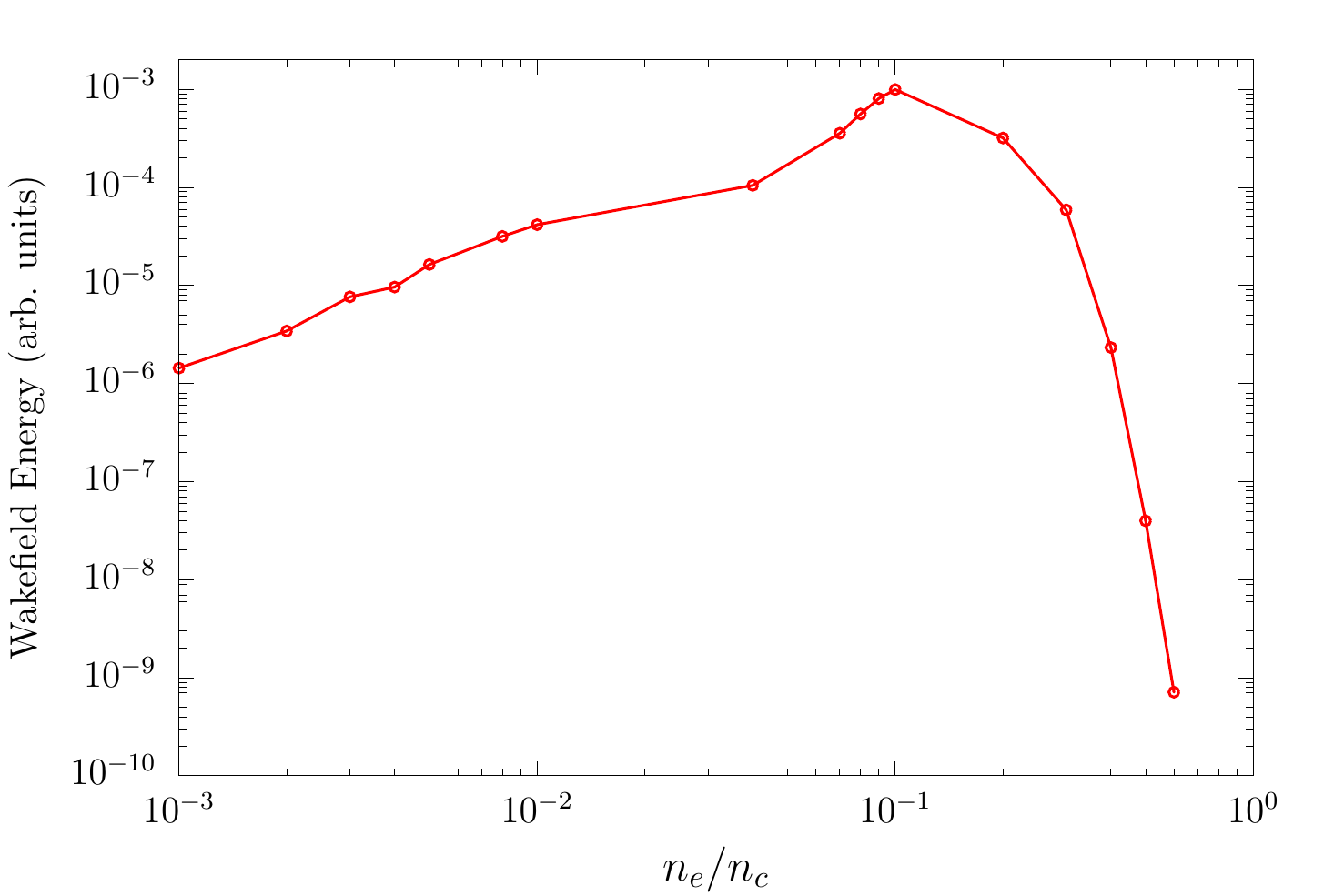}
\end{center}
\caption{The
wakefield energy is calculated as $\protect\int_{z_0}^{z_1} |E_z|^2 dz$.
Here, $z_0$ and $z_1$ are $20\protect\lambda$ and $10\protect\lambda$ behind
the peak of the laser at $t = 263$ for different plasma densities. }
\label{WakeEner}
\end{figure}

As we have seen in Fig \ref{dispersion}(a,b), for the higher plasma
densities wakefield generation will be suppressed due to dispersion, unless
the amplitude is strong enough to keep the pulse short enough, as in Fig \ref%
{dispersion}(c). To avoid  dispersive suppression we will consider pulse
amplitudes firmly in the nonlinear (= non-dispersive) regime and pick $%
a_{0}=0.3$. \ We are interested in the dependence of the wakefield energy as
a function of density (rather than the absolute value), and therefore we
have used the integral $\int_{z_{0}}^{z_{1}}|E_{z}|^{2}dz$ as a measure.
Here $z_{0}$ and $z_{1}$ are $20\lambda $ and $10\lambda $ behind the peak
of the laser computed at $t=263$ for a given plasma density. Note that the
wakefield will divide its energy equally between kinetic energy and
electrostatic field energy, unless we are in an extremely nonlinear regime.
The generated wakefield energy as a function of the plasma density is
presented in Fig. \ref{WakeEner}. Initially the wakefield energy increases
with density, as expected from standard theory of wakefield generation. The
maximum peak of the wakefield energy is reached for a density of the order $%
0.1n_{c}$. However, if we further increase the plasmas density beyond
approximately $n_{e}\sim 0.2n_{c}$, there is rapid drop in the wakefield
energy. The drop in wakefield energy coincides with the formation of
solitary pulses which we label quasi-solitons, as discussed in the previous
sub-section. 

\begin{figure}[b]
\begin{center}
\includegraphics[totalheight=2.2in]{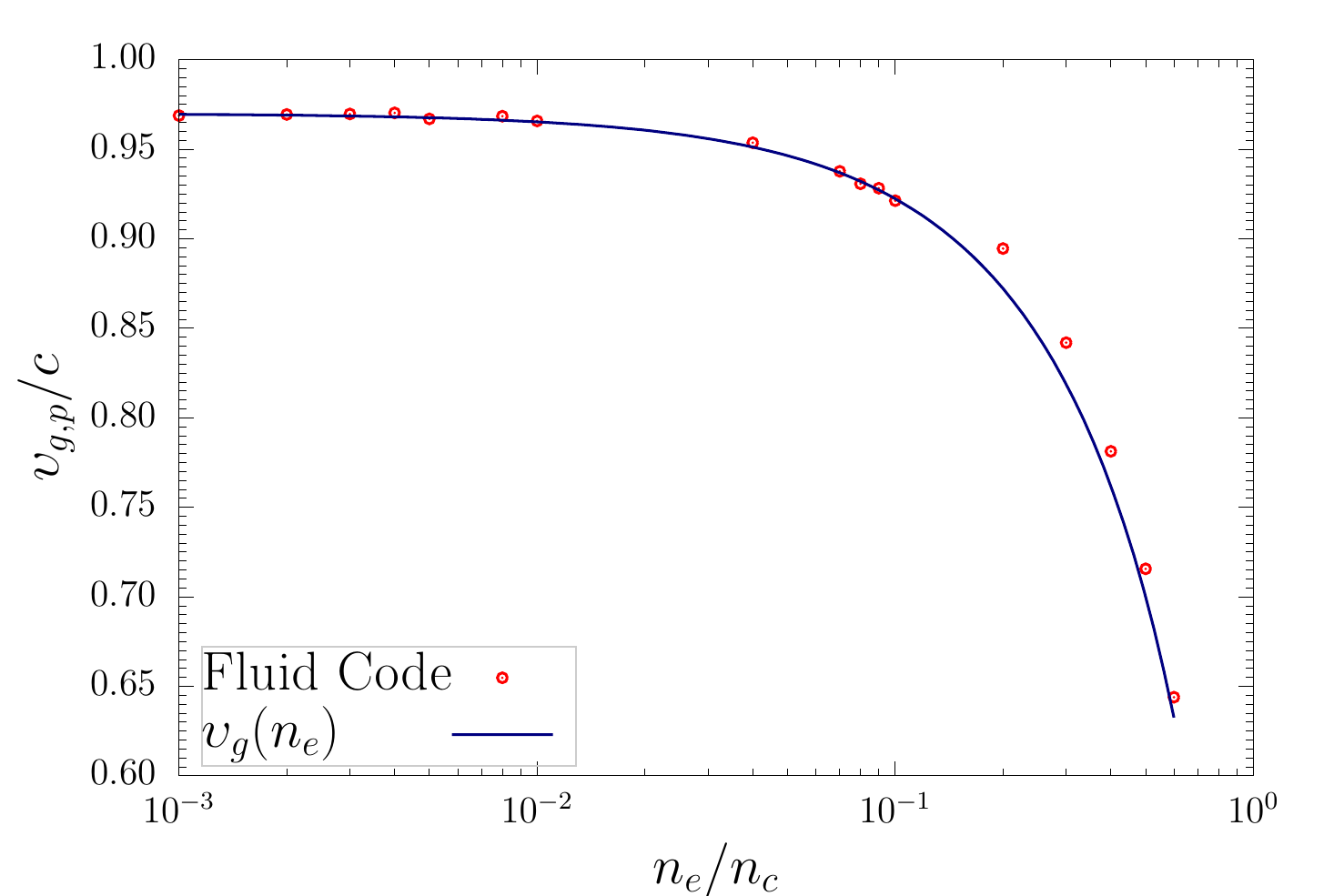}
\end{center}
\caption{A comparison of the group velocity $\upsilon_g$, the line fitted by Eq. \protect\ref{groupvel-2},
and the propagation velocity $\upsilon_p$, the dots computed from the position
of the central peak of the EM field, as a function of plasma density.
Here the case of a 5 cycle EM pulse with $a_0=0.3$ is considered. }
\label{GVelo}
\end{figure}

\begin{figure*}[t]
\begin{center}
\includegraphics[totalheight=3in]{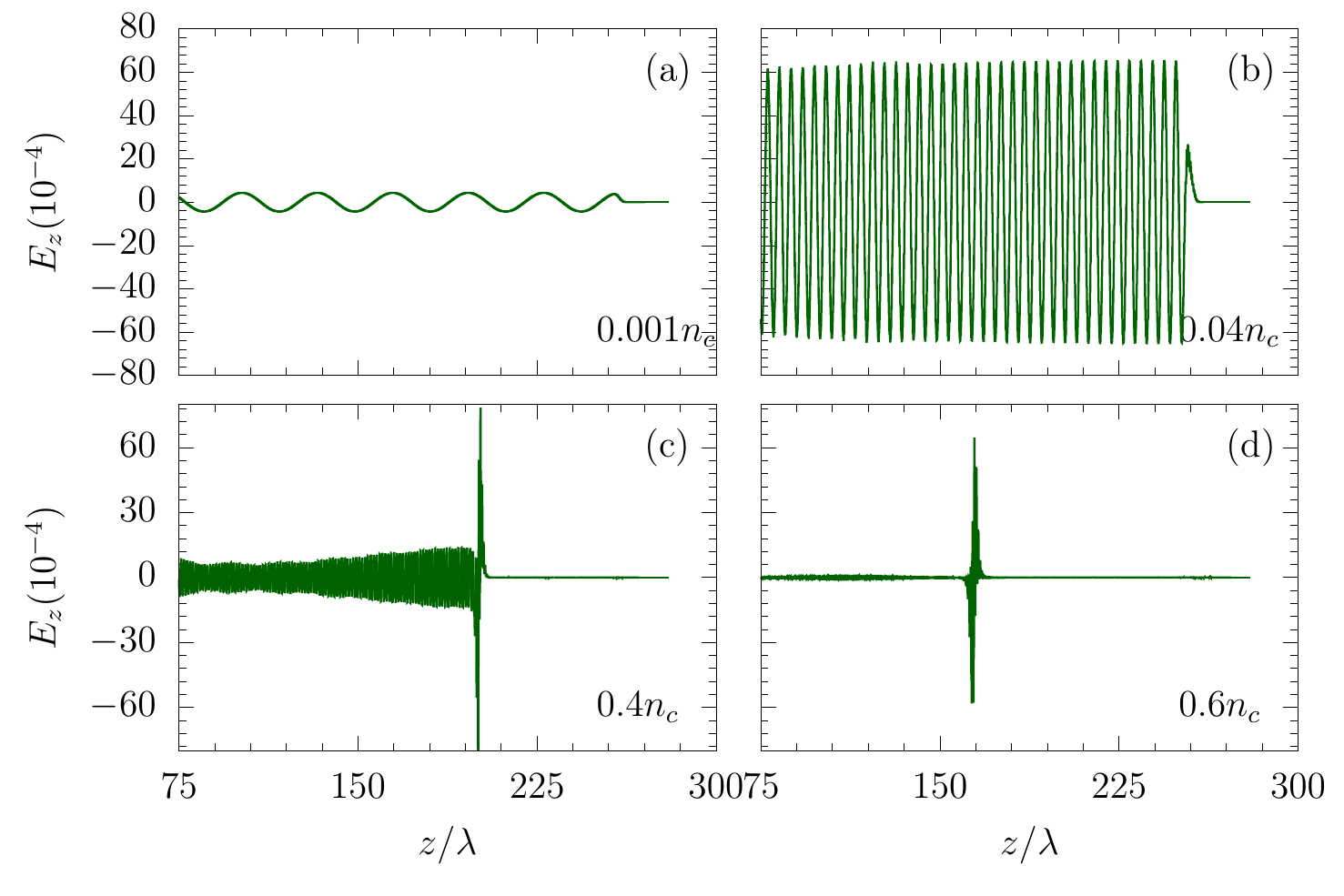}
\end{center}
\caption{PIC simulation: The wakefield for different plasma densities 0.001$%
n_c$ (a), 0.04$n_c$ (b), 0.4$n_c$ (c) and 0.6$n_c$ (d) are presented as
calculated at $t = 265$. The laser parameters are $a_0 = 0.3$, 5 cycle
(FWHM) Gaussian linearly polarized laser pulse, incident on the plasma slab
of $275\protect\lambda$. }
\label{PICWake}
\end{figure*}

\section{Comparison with theory and PIC-simulations}

A first theoretical estimate of the propagation velocity for the EM-pulses
can be found by computing the group velocity including a nonlinear
correction for the relativistic factor induced by the EM field. \ In
linearized theory the group velocity can be calculated as 
\begin{equation}
\upsilon _{g}=c\sqrt{1-\omega _{p}^{2}/\omega ^{2}}=c\sqrt{\varepsilon }
\label{Group-velocity-1}
\end{equation}%
To get a first estimate of the amplitude dependence, we can replace the
unperturbed plasma frequency with the value corrected by the gamma factor
due the transverse motion in the EM-field. This suggests that the expression
from linearized theory can be substituted as $ne^{2}/\varepsilon
_{0}m_{e}\rightarrow ne^{2}/\varepsilon _{0}\gamma m_{e}$ where we let $%
\gamma =\sqrt{1+a_{0}^{2}},$ where we use the normalized peak vector
potential. Using the peak potential overestimates the average gamma factor,
but ignoring the contribution from the longitudinal motion underestimates
it. Hence this may serve as a rough estimate for the propagation velocity.
The expression which we will compare with the results from the numerical
code is now given by 
\begin{equation}
\upsilon _{g}=c\sqrt{1-\frac{n_{e}}{n_{c}\sqrt{1+a_{0}^{2}}}}.
\label{groupvel-2}
\end{equation}%
In spite of the rather crude estimates involved, we see that the propagation
velocity computed by the fluid code (simply based on the
position of the central peak of the EM-field)  agrees very well with the expression (%
\ref{groupvel-2}), as we can see in Fig \ref{GVelo}. Here the propagation
velocity is plotted as a function of density for the case of $a_{0}=0.3$. 

Next we turn to a comparison of the fluid code with PIC-simulations. The 1D
Particle-In-Cell simulation (LPIC++) \cite{lpic} is carried out to study the
effect of plasma densities on the wakefield generation. For the results
presented here the EM field amplitude of the driver is considered to be 0.3
(in dimensionless unit $a_{0}=eE/m\omega c$, with $\omega $ being the
frequency), space and time are taken in units of laser wavelength ($\lambda $%
) and one laser cycle $\tau =\lambda /c$ respectively. The 800 nm laser with
FWHM duration of 5$\tau $ and an initially Gaussian shape propagates through
the plasma along the $z$-direction with electric field along $x$-direction. The plasma of length 275 $\lambda $ is considered having uniform density $n_{0}$.

In Fig. \ref{PICWake}, the spatial profile of longitudinal field $E_{z}$ for
different plasma densities calculated at $t=265$ is presented. It can be
seen that for low plasma densities we have wakefield generation behind the
laser pulse for low density plasmas ($0.001\leq n_{e}\leq 0.04n_{c}$).
However, as the plasma density increases the wakefield amplitude decreases
and eventually we have only a soliton like structure propagating in the
plasma for say $n_{e}=0.6n_{c}$. These results are in good agreement with
the fluid results, as can be seen by comparing with the corresponding
results from the fluid code shown in Fig. \ref{fluidWake}.

\section{Summary and discussion}

In the present paper we have derived a cold, fully relativistic 1D model,
that we have solved numerically. The purpose has been to study the
competition between linear dispersion, nonlinear self-modulation and
wakefield generation for a localized electromagnetic pulse propagating in a
homogeneous non-magnetized plasma. First we have deduced that wakefield
generation for initially short pulses is suppressed by linear dispersion
after a relatively short time, if the plasma density is modest or high (i.e.
of the order $0.1n_{c}$ or higher) unless the normalized amplitude is of the
order $a_{0}\simeq 0.3$. In case we are firmly in the nonlinear regime, the
nonlinear self-modulation keep the pulse short enough to sustain wakefield
generation for a long time - until depletion due to the wakefield generation
becomes significant. When the plasma density is increased further, wakefield
generation is replaced by a formation of quasi-solitons. The central part of
the pulse propagates with very little changes of the pulse shape, but a
small irradiation backward and forward of the low-frequency and
high-frequency part of the spectrum still occur. A key part of the present
study is the scaling with the plasma density of the energy loss of the
EM-pulse due to wakefield generation. For pulses in the nonlinear regime
(not suffering dispersive broadening), we find that the energy loss has a
maximum when the plasma density is of the order $n\simeq 0.1n_{c}$. When the
plasma density is increased beyond $n\simeq 0.2n_{c}$ there is a rapid
decrease of the wakefield energy with the plasma density, signalling the
onset of (quasi) soliton formation. The results produced by the fluid code
has been checked against 1D PIC simulations, and the agreement has been
found to be excellent.

Much of the conclusions in the present paper has a rather general nature, as
the competition between dispersion, wakefield generation and self-modulation
may take place in a magnetized plasma, and also for different types of wave
modes. Qualitatively many of the features of the present study can be
assumed to hold in a more general context, as long as the nonlinear
self-modulation is of focusing type. In particular wakefield generation is
more prominent for short wavepackets (which may be of electromagnetic or
electrostatic nature), and are likely to be suppressed by dispersive
broadening, unless the nonlinear self-modulation prevents this from
happening. Moreover, the energy loss due to wakefield generation is likely
to be suppressed for low and high plasma densities (for a given wave
frequency). Thus the existence of a plasma density that optimizes wakefield
generation is likely a generic feature. To what extent the
conclusions reported here holds also for electromagnetic waves propagating
in a magnetized plasma remains an issue for further research.

\section*{Acknowledgements}

AH acknowledges \emph{Department of Physics, Ume\aa\ University, Sweden} for
the local hospitality and the travel support. AH also acknowledges \emph{Max
Planck Institute for the Physics of Complex Systems, Dresden, Germany} for
the financial support. GB would like to acknowledge financial support by the 
Swedish Research Council, grant number 2016-03806.

 

\begin{thebibliography}{17}%
\makeatletter
\providecommand \@ifxundefined [1]{%
 \@ifx{#1\undefined}
}%
\providecommand \@ifnum [1]{%
 \ifnum #1\expandafter \@firstoftwo
 \else \expandafter \@secondoftwo
 \fi
}%
\providecommand \@ifx [1]{%
 \ifx #1\expandafter \@firstoftwo
 \else \expandafter \@secondoftwo
 \fi
}%
\providecommand \natexlab [1]{#1}%
\providecommand \enquote  [1]{``#1''}%
\providecommand \bibnamefont  [1]{#1}%
\providecommand \bibfnamefont [1]{#1}%
\providecommand \citenamefont [1]{#1}%
\providecommand \href@noop [0]{\@secondoftwo}%
\providecommand \href [0]{\begingroup \@sanitize@url \@href}%
\providecommand \@href[1]{\@@startlink{#1}\@@href}%
\providecommand \@@href[1]{\endgroup#1\@@endlink}%
\providecommand \@sanitize@url [0]{\catcode `\\12\catcode `\$12\catcode
  `\&12\catcode `\#12\catcode `\^12\catcode `\_12\catcode `\%12\relax}%
\providecommand \@@startlink[1]{}%
\providecommand \@@endlink[0]{}%
\providecommand \url  [0]{\begingroup\@sanitize@url \@url }%
\providecommand \@url [1]{\endgroup\@href {#1}{\urlprefix }}%
\providecommand \urlprefix  [0]{URL }%
\providecommand \Eprint [0]{\href }%
\providecommand \doibase [0]{http://dx.doi.org/}%
\providecommand \selectlanguage [0]{\@gobble}%
\providecommand \bibinfo  [0]{\@secondoftwo}%
\providecommand \bibfield  [0]{\@secondoftwo}%
\providecommand \translation [1]{[#1]}%
\providecommand \BibitemOpen [0]{}%
\providecommand \bibitemStop [0]{}%
\providecommand \bibitemNoStop [0]{.\EOS\space}%
\providecommand \EOS [0]{\spacefactor3000\relax}%
\providecommand \BibitemShut  [1]{\csname bibitem#1\endcsname}%
\let\auto@bib@innerbib\@empty
\bibitem [{\citenamefont {Tajima}\ and\ \citenamefont
  {Dawson}(1979)}]{Tajima1979_prl}%
  \BibitemOpen
  \bibfield  {author} {\bibinfo {author} {\bibfnamefont {T.}~\bibnamefont
  {Tajima}}\ and\ \bibinfo {author} {\bibfnamefont {J.~M.}\ \bibnamefont
  {Dawson}},\ }\href {\doibase 10.1103/PhysRevLett.43.267} {\bibfield
  {journal} {\bibinfo  {journal} {Phys. Rev. Lett.}\ }\textbf {\bibinfo
  {volume} {43}},\ \bibinfo {pages} {267} (\bibinfo {year} {1979})}\BibitemShut
  {NoStop}%
\bibitem [{\citenamefont {Kneip}\ \emph {et~al.}(2009)\citenamefont {Kneip},
  \citenamefont {Nagel}, \citenamefont {Martins}, \citenamefont {Mangles},
  \citenamefont {Bellei}, \citenamefont {Chekhlov}, \citenamefont {Clarke},
  \citenamefont {Delerue}, \citenamefont {Divall}, \citenamefont {Doucas},
  \citenamefont {Ertel}, \citenamefont {Fiuza}, \citenamefont {Fonseca},
  \citenamefont {Foster}, \citenamefont {Hawkes}, \citenamefont {Hooker},
  \citenamefont {Krushelnick}, \citenamefont {Mori}, \citenamefont {Palmer},
  \citenamefont {Phuoc}, \citenamefont {Rajeev}, \citenamefont {Schreiber},
  \citenamefont {Streeter}, \citenamefont {Urner}, \citenamefont {Vieira},
  \citenamefont {Silva},\ and\ \citenamefont {Najmudin}}]{Kneip2009_prl}%
  \BibitemOpen
  \bibfield  {author} {\bibinfo {author} {\bibfnamefont {S.}~\bibnamefont
  {Kneip}}, \bibinfo {author} {\bibfnamefont {S.~R.}\ \bibnamefont {Nagel}},
  \bibinfo {author} {\bibfnamefont {S.~F.}\ \bibnamefont {Martins}}, \bibinfo
  {author} {\bibfnamefont {S.~P.~D.}\ \bibnamefont {Mangles}}, \bibinfo
  {author} {\bibfnamefont {C.}~\bibnamefont {Bellei}}, \bibinfo {author}
  {\bibfnamefont {O.}~\bibnamefont {Chekhlov}}, \bibinfo {author}
  {\bibfnamefont {R.~J.}\ \bibnamefont {Clarke}}, \bibinfo {author}
  {\bibfnamefont {N.}~\bibnamefont {Delerue}}, \bibinfo {author} {\bibfnamefont
  {E.~J.}\ \bibnamefont {Divall}}, \bibinfo {author} {\bibfnamefont
  {G.}~\bibnamefont {Doucas}}, \bibinfo {author} {\bibfnamefont
  {K.}~\bibnamefont {Ertel}}, \bibinfo {author} {\bibfnamefont
  {F.}~\bibnamefont {Fiuza}}, \bibinfo {author} {\bibfnamefont
  {R.}~\bibnamefont {Fonseca}}, \bibinfo {author} {\bibfnamefont
  {P.}~\bibnamefont {Foster}}, \bibinfo {author} {\bibfnamefont {S.~J.}\
  \bibnamefont {Hawkes}}, \bibinfo {author} {\bibfnamefont {C.~J.}\
  \bibnamefont {Hooker}}, \bibinfo {author} {\bibfnamefont {K.}~\bibnamefont
  {Krushelnick}}, \bibinfo {author} {\bibfnamefont {W.~B.}\ \bibnamefont
  {Mori}}, \bibinfo {author} {\bibfnamefont {C.~A.~J.}\ \bibnamefont {Palmer}},
  \bibinfo {author} {\bibfnamefont {K.~T.}\ \bibnamefont {Phuoc}}, \bibinfo
  {author} {\bibfnamefont {P.~P.}\ \bibnamefont {Rajeev}}, \bibinfo {author}
  {\bibfnamefont {J.}~\bibnamefont {Schreiber}}, \bibinfo {author}
  {\bibfnamefont {M.~J.~V.}\ \bibnamefont {Streeter}}, \bibinfo {author}
  {\bibfnamefont {D.}~\bibnamefont {Urner}}, \bibinfo {author} {\bibfnamefont
  {J.}~\bibnamefont {Vieira}}, \bibinfo {author} {\bibfnamefont {L.~O.}\
  \bibnamefont {Silva}}, \ and\ \bibinfo {author} {\bibfnamefont
  {Z.}~\bibnamefont {Najmudin}},\ }\href {\doibase
  10.1103/PhysRevLett.103.035002} {\bibfield  {journal} {\bibinfo  {journal}
  {Phys. Rev. Lett.}\ }\textbf {\bibinfo {volume} {103}},\ \bibinfo {pages}
  {035002} (\bibinfo {year} {2009})}\BibitemShut {NoStop}%
\bibitem [{\citenamefont {Kim}\ \emph {et~al.}(2013)\citenamefont {Kim},
  \citenamefont {Pae}, \citenamefont {Cha}, \citenamefont {Kim}, \citenamefont
  {Yu}, \citenamefont {Sung}, \citenamefont {Lee}, \citenamefont {Jeong},\ and\
  \citenamefont {Lee}}]{kim2013_prl}%
  \BibitemOpen
  \bibfield  {author} {\bibinfo {author} {\bibfnamefont {H.~T.}\ \bibnamefont
  {Kim}}, \bibinfo {author} {\bibfnamefont {K.~H.}\ \bibnamefont {Pae}},
  \bibinfo {author} {\bibfnamefont {H.~J.}\ \bibnamefont {Cha}}, \bibinfo
  {author} {\bibfnamefont {I.~J.}\ \bibnamefont {Kim}}, \bibinfo {author}
  {\bibfnamefont {T.~J.}\ \bibnamefont {Yu}}, \bibinfo {author} {\bibfnamefont
  {J.~H.}\ \bibnamefont {Sung}}, \bibinfo {author} {\bibfnamefont {S.~K.}\
  \bibnamefont {Lee}}, \bibinfo {author} {\bibfnamefont {T.~M.}\ \bibnamefont
  {Jeong}}, \ and\ \bibinfo {author} {\bibfnamefont {J.}~\bibnamefont {Lee}},\
  }\href {\doibase 10.1103/PhysRevLett.111.165002} {\bibfield  {journal}
  {\bibinfo  {journal} {Phys. Rev. Lett.}\ }\textbf {\bibinfo {volume} {111}},\
  \bibinfo {pages} {165002} (\bibinfo {year} {2013})}\BibitemShut {NoStop}%
\bibitem [{\citenamefont {Leemans}\ \emph {et~al.}(2014)\citenamefont
  {Leemans}, \citenamefont {Gonsalves}, \citenamefont {Mao}, \citenamefont
  {Nakamura}, \citenamefont {Benedetti}, \citenamefont {Schroeder},
  \citenamefont {T\'oth}, \citenamefont {Daniels}, \citenamefont
  {Mittelberger}, \citenamefont {Bulanov}, \citenamefont {Vay}, \citenamefont
  {Geddes},\ and\ \citenamefont {Esarey}}]{Leemans2014_prl}%
  \BibitemOpen
  \bibfield  {author} {\bibinfo {author} {\bibfnamefont {W.~P.}\ \bibnamefont
  {Leemans}}, \bibinfo {author} {\bibfnamefont {A.~J.}\ \bibnamefont
  {Gonsalves}}, \bibinfo {author} {\bibfnamefont {H.-S.}\ \bibnamefont {Mao}},
  \bibinfo {author} {\bibfnamefont {K.}~\bibnamefont {Nakamura}}, \bibinfo
  {author} {\bibfnamefont {C.}~\bibnamefont {Benedetti}}, \bibinfo {author}
  {\bibfnamefont {C.~B.}\ \bibnamefont {Schroeder}}, \bibinfo {author}
  {\bibfnamefont {C.}~\bibnamefont {T\'oth}}, \bibinfo {author} {\bibfnamefont
  {J.}~\bibnamefont {Daniels}}, \bibinfo {author} {\bibfnamefont {D.~E.}\
  \bibnamefont {Mittelberger}}, \bibinfo {author} {\bibfnamefont {S.~S.}\
  \bibnamefont {Bulanov}}, \bibinfo {author} {\bibfnamefont {J.-L.}\
  \bibnamefont {Vay}}, \bibinfo {author} {\bibfnamefont {C.~G.~R.}\
  \bibnamefont {Geddes}}, \ and\ \bibinfo {author} {\bibfnamefont
  {E.}~\bibnamefont {Esarey}},\ }\href {\doibase
  10.1103/PhysRevLett.113.245002} {\bibfield  {journal} {\bibinfo  {journal}
  {Phys. Rev. Lett.}\ }\textbf {\bibinfo {volume} {113}},\ \bibinfo {pages}
  {245002} (\bibinfo {year} {2014})}\BibitemShut {NoStop}%
\bibitem [{\citenamefont {Shen}\ \emph {et~al.}(2007)\citenamefont {Shen},
  \citenamefont {Li}, \citenamefont {Yu},\ and\ \citenamefont
  {Cary}}]{PhysRevE.76.055402}%
  \BibitemOpen
  \bibfield  {author} {\bibinfo {author} {\bibfnamefont {B.}~\bibnamefont
  {Shen}}, \bibinfo {author} {\bibfnamefont {Y.}~\bibnamefont {Li}}, \bibinfo
  {author} {\bibfnamefont {M.~Y.}\ \bibnamefont {Yu}}, \ and\ \bibinfo {author}
  {\bibfnamefont {J.}~\bibnamefont {Cary}},\ }\href {\doibase
  10.1103/PhysRevE.76.055402} {\bibfield  {journal} {\bibinfo  {journal} {Phys.
  Rev. E}\ }\textbf {\bibinfo {volume} {76}},\ \bibinfo {pages} {055402}
  (\bibinfo {year} {2007})}\BibitemShut {NoStop}%
\bibitem [{\citenamefont {Hidding}\ \emph {et~al.}(2010)\citenamefont
  {Hidding}, \citenamefont {K\"onigstein}, \citenamefont {Osterholz},
  \citenamefont {Karsch}, \citenamefont {Willi},\ and\ \citenamefont
  {Pretzler}}]{PhysRevLett.104.195002}%
  \BibitemOpen
  \bibfield  {author} {\bibinfo {author} {\bibfnamefont {B.}~\bibnamefont
  {Hidding}}, \bibinfo {author} {\bibfnamefont {T.}~\bibnamefont
  {K\"onigstein}}, \bibinfo {author} {\bibfnamefont {J.}~\bibnamefont
  {Osterholz}}, \bibinfo {author} {\bibfnamefont {S.}~\bibnamefont {Karsch}},
  \bibinfo {author} {\bibfnamefont {O.}~\bibnamefont {Willi}}, \ and\ \bibinfo
  {author} {\bibfnamefont {G.}~\bibnamefont {Pretzler}},\ }\href {\doibase
  10.1103/PhysRevLett.104.195002} {\bibfield  {journal} {\bibinfo  {journal}
  {Phys. Rev. Lett.}\ }\textbf {\bibinfo {volume} {104}},\ \bibinfo {pages}
  {195002} (\bibinfo {year} {2010})}\BibitemShut {NoStop}%
\bibitem [{\citenamefont {Filip}\ \emph {et~al.}(2004)\citenamefont {Filip},
  \citenamefont {Narang}, \citenamefont {Ya.~Tochitsky}, \citenamefont
  {Clayton}, \citenamefont {Musumeci}, \citenamefont {Yoder}, \citenamefont
  {Marsh}, \citenamefont {Rosenzweig}, \citenamefont {Pellegrini},\ and\
  \citenamefont {Joshi}}]{PhysRevE.69.026404}%
  \BibitemOpen
  \bibfield  {author} {\bibinfo {author} {\bibfnamefont {C.~V.}\ \bibnamefont
  {Filip}}, \bibinfo {author} {\bibfnamefont {R.}~\bibnamefont {Narang}},
  \bibinfo {author} {\bibfnamefont {S.}~\bibnamefont {Ya.~Tochitsky}}, \bibinfo
  {author} {\bibfnamefont {C.~E.}\ \bibnamefont {Clayton}}, \bibinfo {author}
  {\bibfnamefont {P.}~\bibnamefont {Musumeci}}, \bibinfo {author}
  {\bibfnamefont {R.~B.}\ \bibnamefont {Yoder}}, \bibinfo {author}
  {\bibfnamefont {K.~A.}\ \bibnamefont {Marsh}}, \bibinfo {author}
  {\bibfnamefont {J.~B.}\ \bibnamefont {Rosenzweig}}, \bibinfo {author}
  {\bibfnamefont {C.}~\bibnamefont {Pellegrini}}, \ and\ \bibinfo {author}
  {\bibfnamefont {C.}~\bibnamefont {Joshi}},\ }\href {\doibase
  10.1103/PhysRevE.69.026404} {\bibfield  {journal} {\bibinfo  {journal} {Phys.
  Rev. E}\ }\textbf {\bibinfo {volume} {69}},\ \bibinfo {pages} {026404}
  (\bibinfo {year} {2004})}\BibitemShut {NoStop}%
\bibitem [{\citenamefont {Trines}(2009)}]{PhysRevE.79.056406}%
  \BibitemOpen
  \bibfield  {author} {\bibinfo {author} {\bibfnamefont {R.~M. G.~M.}\
  \bibnamefont {Trines}},\ }\href {\doibase 10.1103/PhysRevE.79.056406}
  {\bibfield  {journal} {\bibinfo  {journal} {Phys. Rev. E}\ }\textbf {\bibinfo
  {volume} {79}},\ \bibinfo {pages} {056406} (\bibinfo {year}
  {2009})}\BibitemShut {NoStop}%
\bibitem [{\citenamefont {Hafz}\ \emph {et~al.}(2006)\citenamefont {Hafz},
  \citenamefont {Hur}, \citenamefont {Kim}, \citenamefont {Kim}, \citenamefont
  {Ko},\ and\ \citenamefont {Suk}}]{PhysRevE.73.016405}%
  \BibitemOpen
  \bibfield  {author} {\bibinfo {author} {\bibfnamefont {N.}~\bibnamefont
  {Hafz}}, \bibinfo {author} {\bibfnamefont {M.~S.}\ \bibnamefont {Hur}},
  \bibinfo {author} {\bibfnamefont {G.~H.}\ \bibnamefont {Kim}}, \bibinfo
  {author} {\bibfnamefont {C.}~\bibnamefont {Kim}}, \bibinfo {author}
  {\bibfnamefont {I.~S.}\ \bibnamefont {Ko}}, \ and\ \bibinfo {author}
  {\bibfnamefont {H.}~\bibnamefont {Suk}},\ }\href {\doibase
  10.1103/PhysRevE.73.016405} {\bibfield  {journal} {\bibinfo  {journal} {Phys.
  Rev. E}\ }\textbf {\bibinfo {volume} {73}},\ \bibinfo {pages} {016405}
  (\bibinfo {year} {2006})}\BibitemShut {NoStop}%
\bibitem [{\citenamefont {Holkundkar}, \citenamefont {Brodin},\ and\
  \citenamefont {Marklund}(2011)}]{PhysRevE.84.036409}%
  \BibitemOpen
  \bibfield  {author} {\bibinfo {author} {\bibfnamefont {A.}~\bibnamefont
  {Holkundkar}}, \bibinfo {author} {\bibfnamefont {G.}~\bibnamefont {Brodin}},
  \ and\ \bibinfo {author} {\bibfnamefont {M.}~\bibnamefont {Marklund}},\
  }\href {\doibase 10.1103/PhysRevE.84.036409} {\bibfield  {journal} {\bibinfo
  {journal} {Phys. Rev. E}\ }\textbf {\bibinfo {volume} {84}},\ \bibinfo
  {pages} {036409} (\bibinfo {year} {2011})}\BibitemShut {NoStop}%
\bibitem [{\citenamefont {Pathak}\ \emph {et~al.}(2012)\citenamefont {Pathak},
  \citenamefont {Vieira}, \citenamefont {Fonseca},\ and\ \citenamefont
  {Silva}}]{1367-2630-14-2-023057}%
  \BibitemOpen
  \bibfield  {author} {\bibinfo {author} {\bibfnamefont {V.~B.}\ \bibnamefont
  {Pathak}}, \bibinfo {author} {\bibfnamefont {J.}~\bibnamefont {Vieira}},
  \bibinfo {author} {\bibfnamefont {R.~A.}\ \bibnamefont {Fonseca}}, \ and\
  \bibinfo {author} {\bibfnamefont {L.~O.}\ \bibnamefont {Silva}},\ }\href
  {http://stacks.iop.org/1367-2630/14/i=2/a=023057} {\bibfield  {journal}
  {\bibinfo  {journal} {New Journal of Physics}\ }\textbf {\bibinfo {volume}
  {14}},\ \bibinfo {pages} {023057} (\bibinfo {year} {2012})}\BibitemShut
  {NoStop}%
\bibitem [{\citenamefont {Pukhov}\ and\ \citenamefont
  {Kostyukov}(2008)}]{PhysRevE.77.025401}%
  \BibitemOpen
  \bibfield  {author} {\bibinfo {author} {\bibfnamefont {A.}~\bibnamefont
  {Pukhov}}\ and\ \bibinfo {author} {\bibfnamefont {I.}~\bibnamefont
  {Kostyukov}},\ }\href {\doibase 10.1103/PhysRevE.77.025401} {\bibfield
  {journal} {\bibinfo  {journal} {Phys. Rev. E}\ }\textbf {\bibinfo {volume}
  {77}},\ \bibinfo {pages} {025401} (\bibinfo {year} {2008})}\BibitemShut
  {NoStop}%
\bibitem [{\citenamefont {Yu}, \citenamefont {Shukla},\ and\ \citenamefont
  {Tsintsadze}(1982)}]{Yu1982_PF}%
  \BibitemOpen
  \bibfield  {author} {\bibinfo {author} {\bibfnamefont {M.~Y.}\ \bibnamefont
  {Yu}}, \bibinfo {author} {\bibfnamefont {P.~K.}\ \bibnamefont {Shukla}}, \
  and\ \bibinfo {author} {\bibfnamefont {N.~L.}\ \bibnamefont {Tsintsadze}},\
  }\href {\doibase 10.1063/1.863836} {\bibfield  {journal} {\bibinfo  {journal}
  {Phys. Fluids}\ }\textbf {\bibinfo {volume} {25}},\ \bibinfo {pages} {1049}
  (\bibinfo {year} {1982})}\BibitemShut {NoStop}%
\bibitem [{\citenamefont {Naumova}\ \emph {et~al.}(2001)\citenamefont
  {Naumova}, \citenamefont {Bulanov}, \citenamefont {Esirkepov}, \citenamefont
  {Farina}, \citenamefont {Nishihara}, \citenamefont {Pegoraro}, \citenamefont
  {Ruhl},\ and\ \citenamefont {Sakharov}}]{NaumovaPRL_2001}%
  \BibitemOpen
  \bibfield  {author} {\bibinfo {author} {\bibfnamefont {N.~M.}\ \bibnamefont
  {Naumova}}, \bibinfo {author} {\bibfnamefont {S.~V.}\ \bibnamefont
  {Bulanov}}, \bibinfo {author} {\bibfnamefont {T.~Z.}\ \bibnamefont
  {Esirkepov}}, \bibinfo {author} {\bibfnamefont {D.}~\bibnamefont {Farina}},
  \bibinfo {author} {\bibfnamefont {K.}~\bibnamefont {Nishihara}}, \bibinfo
  {author} {\bibfnamefont {F.}~\bibnamefont {Pegoraro}}, \bibinfo {author}
  {\bibfnamefont {H.}~\bibnamefont {Ruhl}}, \ and\ \bibinfo {author}
  {\bibfnamefont {A.~S.}\ \bibnamefont {Sakharov}},\ }\href {\doibase
  10.1103/PhysRevLett.87.185004} {\bibfield  {journal} {\bibinfo  {journal}
  {Phys. Rev. Lett.}\ }\textbf {\bibinfo {volume} {87}},\ \bibinfo {pages}
  {185004} (\bibinfo {year} {2001})}\BibitemShut {NoStop}%
\bibitem [{\citenamefont {Esirkepov}\ \emph {et~al.}(2002)\citenamefont
  {Esirkepov}, \citenamefont {Nishihara}, \citenamefont {Bulanov},\ and\
  \citenamefont {Pegoraro}}]{EsirkepovPRL_2002}%
  \BibitemOpen
  \bibfield  {author} {\bibinfo {author} {\bibfnamefont {T.}~\bibnamefont
  {Esirkepov}}, \bibinfo {author} {\bibfnamefont {K.}~\bibnamefont
  {Nishihara}}, \bibinfo {author} {\bibfnamefont {S.~V.}\ \bibnamefont
  {Bulanov}}, \ and\ \bibinfo {author} {\bibfnamefont {F.}~\bibnamefont
  {Pegoraro}},\ }\href {\doibase 10.1103/PhysRevLett.89.275002} {\bibfield
  {journal} {\bibinfo  {journal} {Phys. Rev. Lett.}\ }\textbf {\bibinfo
  {volume} {89}},\ \bibinfo {pages} {275002} (\bibinfo {year}
  {2002})}\BibitemShut {NoStop}%
\bibitem [{\citenamefont {Jovanovi\'{c}}\ \emph {et~al.}(2015)\citenamefont
  {Jovanovi\'{c}}, \citenamefont {Fedele}, \citenamefont {Beli\'{c}},\ and\
  \citenamefont {Nicola}}]{Jovanovic2015_POP}%
  \BibitemOpen
  \bibfield  {author} {\bibinfo {author} {\bibfnamefont {D.}~\bibnamefont
  {Jovanovi\'{c}}}, \bibinfo {author} {\bibfnamefont {R.}~\bibnamefont
  {Fedele}}, \bibinfo {author} {\bibfnamefont {M.}~\bibnamefont {Beli\'{c}}}, \
  and\ \bibinfo {author} {\bibfnamefont {S.~D.}\ \bibnamefont {Nicola}},\
  }\href {\doibase 10.1063/1.4916909} {\bibfield  {journal} {\bibinfo
  {journal} {Phys. Plasmas}\ }\textbf {\bibinfo {volume} {22}},\ \bibinfo
  {pages} {043110} (\bibinfo {year} {2015})}\BibitemShut {NoStop}%
\bibitem [{\citenamefont {Lichters}, \citenamefont {Pfund},\ and\ \citenamefont
  {Meyer-Ter-Vehn}(1997)}]{lpic}%
  \BibitemOpen
  \bibfield  {author} {\bibinfo {author} {\bibfnamefont {R.}~\bibnamefont
  {Lichters}}, \bibinfo {author} {\bibfnamefont {R.~E.~W.}\ \bibnamefont
  {Pfund}}, \ and\ \bibinfo {author} {\bibfnamefont {J.}~\bibnamefont
  {Meyer-Ter-Vehn}},\ }\href {http://www.lichters.net/download.html} {\bibfield
   {journal} {\bibinfo  {journal} {MPQ Report}\ }\textbf {\bibinfo {volume}
  {225}} (\bibinfo {year} {1997})}\BibitemShut {NoStop}%
\end{thebibliography}

%

\end{document}